\documentclass{amsart}

\theoremstyle{definition}

\theoremstyle{remark}

\numberwithin{equation}{section}

\begin{document}

\title{Estimation of within-study covariances in multivariate meta-analysis}

\author{Xiaohuan Xue}
\address{Department of Mathematics and Statistics, Wake Forest University, Winston-Salem, North Carolina 27109}
\curraddr{Department of Psychology, University of California, Los Angeles, California 90095}
\email{xhxue@g.ucla.edu}

%\author{Author Two}
%\address{Mathematical Research Section, School of Mathematical Sciences,
%Australian National University, Canberra ACT 2601, Australia}

%    General info

%\date{January 1, 2001 and, in revised form, June 22, 2001.}



\begin{abstract}
Multivariate meta-analysis can be adapted to a wide range of situations for multiple outcomes and multiple treatment groups when combining studies together. The within-study correlation between effect sizes is often assumed known in multivariate meta-analysis while it is not always known practically. In this paper, we propose a generic method to approximate the within-study covariance for effect sizes in multivariate meta-analysis and apply this method to the scenarios with multiple outcomes and one outcome with multiple treatment groups respectively. 
\end{abstract}

\maketitle

\section{Introduction}

Meta-analysis combines information from multiple existing studies to have better knowledge of the overall effect sizes, which makes it widely used in many research areas \cite{Hedges Text, Card, Intro}. Meta-analysis can give more accurate information about the effect size of interest by integrating the results of multiple studies and thus expanding sample size without conducting studies. The logic of meta-analysis gives rise to the two levels in meta-analysis, within-study level considers the model of a specific study and between-study level models the variability between studies.

According to variability assumptions of within study and between study, there are three types of meta-analysis models\cite{Hedges Text, Card, Intro}: fixed-effects models, random-effects models, and mixed-effects models. Fixed-effects models assume that there is no between-study level variation for effect size and effect size is the same for every study, while random-effects models assume that the effect size is also a  variable for different studies, which is characterized by between-study heterogeneity, the variance of the distribution of study specific true effect sizes. The mixed-effects model, also referred to as meta-regression, adds more predictors for with-in study level modeling compared to random-effects models. In practice, we often have the reported within-study information such as effect sizes and sample sizes and we need to infer the between-study information such as overall effect size and between-study heterogeneity.

For the case of univariate meta-analysis, there is one outcome for one treatment and one control group in each study. However, there are often more than one outcome or more than one treatment groups in practice especially with-in study correlation is not zero, which makes multivariate meta-analysis necessary. Under those situations, multivariate meta-analysis often improve the precision of estimating between-study level information.

There are methods like MLE and REMLE to give the overall effect sizes and between-study heterogeneity, but they all assume the within-study variance and covariance for effect sizes are known while it is not the case in practice. In this paper, we propose a novel method to estimate the within-study variance and covariance matrix for effect sizes and we compute the variance-covariance matrix for two scenarios: two outcomes for one treatment and one control group in each study and one outcome for multiple treatment groups and one control group in each study. 

\section{A novel approach of covariance for standardized mean differences of two outcomes one treatment one control}

For the case of multiple outcomes for one treatment group and one control group in each study, there is an existing method \cite{Wei} being applied to approximate the within-study correlation between standardized mean differences by applying delta method twice and a result from moment generating function \cite{White}. We now propose a method that applies delta method once and it can also work for more complicated effect sizes. For standard mean difference, it can be expressed as a function in terms of mean difference and variance, which makes estimating the with-in study covariance of standard mean differences much more easier since we only need to take derivatives here.

By constructing the functions of known statistics in delta method, we can have the approximation of any possible effect sizes.

We start with the case of two outcomes discussed in Wei's paper \cite{Wei}. At the with-in study level, let $y_{ji}^{t}$ and $y_{ji}^{c}$ be the $ith$ individual of treatment and outcome for outcome $j$ respectively. Let $n_j^t$ and $n_j^c$ be the number of participants in treatment group and control group for outcome $j$ respectively. For fixed $i$, we assume $y_{ji}^{t}$ and $y_{ji}^{c}$ are normally distributed, that is,

 \begin{equation}
  y_{ji}^{t}\sim N(\mu_j^{t},\sigma_j^2)
 \end{equation}

  \begin{equation}
  y_{ji}^{c}\sim N(\mu_j^{c},\sigma_j^2)
 \end{equation}

Standardized mean difference is our interested effect size and Hedges' estimator $g$ is an unbiased estimator of it. Hedges' $g$ is given by the sample mean difference divided by a multiple of the pooled sample standard deviation as following

\begin{equation}
g_j=J(v_j)\cdot\frac{\overline{y}_j^{t}-\overline{y}_j^{c}}{s_{jp}}, J(v_j)=\frac{\Gamma(\frac{v_j}{2})}{\sqrt{\frac{v_j}{2}}\Gamma(\frac{v_j-1}{2})}
\end{equation}
where $s_{jp}$ is the pooled sample standard deviation for outcome $j$
\begin{equation}
s_{jp}=\sqrt{\frac{(n_j^{t}-1)(s_j^{t})^2+(n_j^{c}-1)(s_j^{c})^2}{n_j^{t}+n_j^{c}-2}}
\end{equation}

To find out $\mathrm{Cov}(g_j,g_{j'})$, we use delta method to get an approximation. We treat $g_j$ as a function of $\overline{y}_j^{t}-\overline{y}_j^{c}$ and $s^2_{jp}$, which reduces the complexity. Let
\begin{equation}
X=(\overline{y}_j^{t}-\overline{y}_j^{c},s^2_{jp},\overline{y}_{j'}^{t}-\overline{y}_{j'}^{c},s^2_{j'p})=(X_1,X_2,X_3,X_4)
\end{equation}
and
 \begin{equation}
 \begin{cases}
 f_1(X)=J(v_j)\cdot\frac{\overline{y}_j^{t}-\overline{y}_j^{c}}{s_{jp}}=J(v_j)\frac{X_1}{\sqrt{X_2}}\\
 f_2(X)=J(v_{j'})\cdot\frac{\overline{y}_{j'}^{t}-\overline{y}_{j'}^{c}}{s_{j'p}}=J(v_{j'})\frac{X_3}{\sqrt{X_4}}
 \end{cases}
\end{equation}
then we have
\begin{equation}
\mathrm{Cov}(g_j,g_{j'})=\mathrm{Cov}(J(v_j)\cdot\frac{\overline{y}_j^{t}-\overline{y}_j^{c}}{s_{jp}},J(v_{j'})\cdot\frac{\overline{y}_{j'}^{t}-\overline{y}_{j'}^{c}}{s_{j'p}})=\mathrm{Cov}(f_1(X),f_2(X))
\end{equation}

By delta method,
 \begin{equation}
\mathrm{Var}(f(X))\approx \nabla f^T \mathrm{Var}(X)\nabla f
 \end{equation}
we need to find out $\nabla f$ evaluated at $\mathrm{E}(X)$, by taking partial derivatives we have
\begin{equation}
  \nabla f=
  \begin{bmatrix}
\frac{\partial f_1}{\partial X_1} & \frac{\partial f_2}{\partial X_1} \\
\frac{\partial f_1}{\partial X_2} & \frac{\partial f_2}{\partial X_2} \\
\frac{\partial f_1}{\partial X_3} & \frac{\partial f_2}{\partial X_3}  \\
\frac{\partial f_1}{\partial X_4} & \frac{\partial f_2}{\partial X_4}
\end{bmatrix}
=  \begin{bmatrix}
\frac{J(v_j)}{\sqrt{X_2}} & 0\\
-\frac{J(v_j)X_1}{2\sqrt{X_2^3}} &  0 \\
0 &  \frac{J(v_{j'})}{\sqrt{X_4}} \\
0 & -\frac{X_3J(v_{j'})}{\sqrt{X_4^3}}
\end{bmatrix}
\end{equation}

 Then we need to find out $\mathrm{E}(X)$ and $\mathrm{Var}(X)$ by the assumed distributions given. Since $y_{ji}^{t}$ and $y_{ji}^{c}$ are normally distributed for each $i$, by the properties of normal distributions we have

\begin{equation}
\overline{y}_j^{t} \sim N(\mu_j^{t},\frac{\sigma_j^2}{n_j^{t}}), \overline{y}_j^{c} \sim N(\mu_j^{c},\frac{\sigma_j^2}{n_j^{c}})
\end{equation}

\begin{equation}
(s_j^{t})^2 \cdot\frac{n_j^t-1}{\sigma_j^2}\sim\chi^2_{n_j^t-1}, (s_j^{c})^2 \cdot\frac{n_j^c-1}{\sigma_j^2}\sim\chi^2_{n_j^c-1}
\end{equation}

Thus
\begin{equation}
\overline{y}_j^{t}-\overline{y}_j^{c}  \sim N(\mu_j^{t}-\mu_j^{c},\left(\frac{1}{n_j^{t}}+\frac{1}{n_j^{c}}\right)\sigma_j^2)
\end{equation}

\begin{equation}
s_{jp}^2\cdot\frac{v_j}{\sigma_j^2}\sim\chi^2_{v_j}
\end{equation}
where
\begin{equation}
s_{jp}=\sqrt{\frac{(n_j^{t}-1)(s_j^{t})^2+(n_j^{c}-1)(s_j^{c})^2}{n_j^{t}+n_j^{c}-2}}
\end{equation}
with $v_j=n_j^{t}+n_j^{c}-2$.

Hence the mean and variance of the mean difference and pooled standard deviation are
\begin{equation}
\begin{split}
&\mathrm{E}(MD_j)=\mathrm{E}(\overline{y}_j^{t}-\overline{y}_j^{c})=\mu_j^{t}-\mu_j^{c}\\
&\mathrm{Var}(MD_j)=\mathrm{Var}(\overline{y}_j^{t}-\overline{y}_j^{c})=\left(\frac{1}{n_j^{t}}+\frac{1}{n_j^{c}}\right)\sigma_j^2
\end{split}
\end{equation}

\begin{equation}
\mathrm{E}\left(s_{jp}^2\cdot\frac{v_j}{\sigma_j^2}\right)=v_j
\end{equation}

\begin{equation}
\mathrm{E}\left(s_{jp}^2\right)=\sigma_j^2
\end{equation}

\begin{equation}
\mathrm{Var}\left(s_{jp}^2\cdot\frac{v_j}{\sigma_j^2}\right)=2v_j
\end{equation}

\begin{equation}
\mathrm{Var}\left(s_{jp}^2\right)=2\frac{\sigma_j^4}{v_j}
\end{equation}

Then we can have $\nabla f$ evaluated at $\mathrm{E}(X)$ is
\begin{equation}
  \nabla f=
\begin{bmatrix}
\frac{J(v_j)}{\sqrt{X_2}} & 0\\
-\frac{J(v_j)X_1}{2\sqrt{X_2^3}} &  0 \\
0 &  \frac{J(v_{j'})}{\sqrt{X_4}} \\
0 & -\frac{J(v_{j'})X_3}{2\sqrt{X_4^3}}
\end{bmatrix}
=
\begin{bmatrix}
\frac{J(v_{j})}{\sigma_j} & 0\\
-\frac{J(v_{j})(\mu_j^{t}-\mu_j^{c})}{2\sigma_j^3} &  0 \\
0 &  \frac{J(v_{j'})}{\sigma_{j'}} \\
0 & -\frac{J(v_{j'})(\mu_{j'}^{t}-\mu_{j'}^{c})}{2\sigma_{j'}^3}
\end{bmatrix}
\end{equation}

$\mathrm{Var}(X)$ can also be obtained by knowing all its entries.

\begin{equation}
\begin{split}
\mathrm{Var}(X)&=
\begin{bmatrix}
\mathrm{Cov}(X_1,X_1) & \mathrm{Cov}(X_1,X_2) &\mathrm{Cov}(X_1,X_3) &\mathrm{Cov}(X_1,X_4)\\
\mathrm{Cov}(X_1,X_2) & \mathrm{Cov}(X_2,X_2) &\mathrm{Cov}(X_2,X_3) &\mathrm{Cov}(X_2,X_4)\\
\mathrm{Cov}(X_1,X_3) & \mathrm{Cov}(X_2,X_3) &\mathrm{Cov}(X_3,X_3) &\mathrm{Cov}(X_3,X_4)\\
\mathrm{Cov}(X_1,X_4) & \mathrm{Cov}(X_2,X_4) &\mathrm{Cov}(X_3,X_4) &\mathrm{Cov}(X_4,X_4)
\end{bmatrix}\\
&=
\begin{bmatrix}
\left(\frac{1}{n_j^{t}}+\frac{1}{n_j^{c}}\right)\sigma_j^2 & 0 &\mathrm{Cov}(X_1,X_3) &0 \\
0 & 2\frac{\sigma_j^4}{v_j} & 0&\mathrm{Cov}(X_2,X_4)\\
\mathrm{Cov}(X_1,X_3)& 0 & \left(\frac{1}{n_{j'}^{t}}+\frac{1}{n_{j'}^{c}}\right)\sigma_{j'}^2 & 0 \\
0 & \mathrm{Cov}(X_2,X_4) & 0& 2\frac{\sigma_{j'}^4}{v_{j'}}\\
\end{bmatrix}
\end{split}
\end{equation}
$\mathrm{Cov}(X_1,X_3)$ and $\mathrm{Cov}(X_2,X_4)$ are given in Wei's paper \cite{Wei} as following
\begin{equation}
  \mathrm{Cov}(X_1,X_3)=\rho_{jj'}\left(\frac{n_{jj'}^t}{n_j^tn_{j'}^t}+\frac{n_{jj'}^c}{n_j^cn_{j'}^c}\right)\sigma_j\sigma_{j'}
\end{equation}

\begin{equation}
  \mathrm{Cov}(X_2,X_4)=\rho^2_{jj'}\sigma^2_j\sigma^2_{j'}k_{jj'}
\end{equation}

By delta method

 \begin{equation}
\mathrm{Var}(f(X))\approx \nabla f^T  \mathrm{Var}(X)\nabla f
 \end{equation}
and from the derivations above,

\begin{equation}
\begin{split}
&\nabla f^T  \mathrm{Var}(X)\nabla f\\
&=
\begin{bmatrix}
\left(\frac{\partial f_1}{\partial X_1}\right)^2\mathrm{Var}(X_1)+\left(\frac{\partial f_1}{\partial X_2}\right)^2\mathrm{Var}(X_2) & \frac{\partial f_1}{\partial X_1}\frac{\partial f_2}{\partial X_3}\mathrm{Cov}(X_1,X_3) + \frac{\partial f_1}{\partial X_2}\frac{\partial f_2}{\partial X_4}\mathrm{Cov}(X_2,X_4)\\
\frac{\partial f_1}{\partial X_1}\frac{\partial f_2}{\partial X_3}\mathrm{Cov}(X_1,X_3) + \frac{\partial f_1}{\partial X_2}\frac{\partial f_2}{\partial X_4}\mathrm{Cov}(X_2,X_4) & \left(\frac{\partial f_2}{\partial X_3}\right)^2\mathrm{Var}(X_3)+\left(\frac{\partial f_2}{\partial X_4}\right)^2\mathrm{Var}(X_4)
\end{bmatrix}\\
&=
\begin{bmatrix}
\left(\frac{1}{n_j^{t}}+\frac{1}{n_j^{c}}+\frac{(\mu_j^{t}-\mu_j^{c})^2}{2v_jJ^2(v_j)\sigma_j^2}\right)J^2(v_j) & \alpha\\
\alpha
 & \left(\frac{1}{n_{j'}^{t}}+\frac{1}{n_{j'}^{c}}+\frac{(\mu_{j'}^{t}-\mu_{j'}^{c})^2}{2v_{j'}\sigma_{j'}^2}\right)J^2(v_{j'})
\end{bmatrix}
\end{split}
\end{equation}
where
\begin{equation}
\alpha=J(v_j)J(v_{j'})\left(\frac{\mathrm{Cov}(X_1,X_3)}{\sigma_j\sigma_{j'}} + \frac{(\mu_j^{t}-\mu_j^{c})(\mu_{j'}^{t}-\mu_{j'}^{c})}{4\sigma_j^3\sigma_{j'}^3}\mathrm{Cov}(X_2,X_4)\right)
\end{equation}

Since by definition,
 \begin{equation}
 \mathrm{Var}(f(X))=
 \begin{pmatrix}
 \mathrm{Cov}(f_1(X),f_1(X)) & \mathrm{Cov}(f_1(X),f_2(X))  \\
 \mathrm{Cov}(f_2(X),f_1(X)) & \mathrm{Cov}(f_2(X),f_2(X))
\end{pmatrix}
 \end{equation}

By comparing each entries of the matrices $\nabla f^T  \mathrm{Var}(X)\nabla f$ and $\mathrm{Var}(f(X))$, we have
\begin{equation}
\begin{split}
\mathrm{Cov}(g_j,g_{j'})&=\mathrm{Cov}(f_1(X),f_2(X))\\
&\approx \rho_{jj'}J(v_j)J(v_{j'})\left(\frac{n_{jj'}^t}{n_j^tn_{j'}^t}+\frac{n_{jj'}^c}{n_j^cn_{j'}^c}\right)+ \rho^2_{jj'}k_{jj'}J(v_j)J(v_{j'})\frac{(\mu_j^{t}-\mu_j^{c})(\mu_{j'}^{t}-\mu_{j'}^{c})}{4\sigma_j\sigma_{j'}}\\
&= \rho_{jj'}J(v_j)J(v_{j'})\left(\frac{n_{jj'}^t}{n_j^tn_{j'}^t}+\frac{n_{jj'}^c}{n_j^cn_{j'}^c}+\rho_{jj'}k_{jj'}\frac{(\mu_j^{t}-\mu_j^{c})(\mu_{j'}^{t}-\mu_{j'}^{c})}{4\sigma_j\sigma_{j'}}\right)
\end{split}
\end{equation}

\section{Meta-Analysis for One Outcome and Multiple Treatments}

In this section we consider the scenario when there are multiple treatments for one outcome in each study. First we give the covariance between mean differences, then we estimate the covariance between standard mean differences by Wei's method and our novel method respectively.

\subsection{Covariance between mean differences}
We start with the within-study level, individual $i$ of treatment $k$ for study $j$ is denoted as $y_{ji}^{tk}$. If individual $i$ is in the control group of study $j$, it is denoted as $y_{ji}^{c}$. First we study the covariance of mean differences of treatments and control within one study, which is independent of distributions. If we assume the treatment groups and control group are independent, we have

 \begin{equation}
 \begin{split}
   &\mathrm{Cov}(MD_j^k,MD_j^{k'})\\
 =&\mathrm{Cov}(\overline{y}_j^{tk}-\overline{y}_j^{c},\overline{y}_j^{tk'}-\overline{y}_j^{c})\\
 =&\mathrm{Cov}(\frac{1}{n_j^{tk}}\sum_iy_{ji}^{tk}-\frac{1}{n_j^{c}}\sum_iy_{ji}^{c},\frac{1}{n_j^{tk'}}\sum_{i'}y_{ji'}^{tk'}-\frac{1}{n_j^{c}}\sum_{i'}y_{ji'}^{c})\\
 =&\frac{1}{n_j^{tk}n_j^{tk'}}\mathrm{Cov}(\sum_iy_{ji}^{tk},\sum_{i'}y_{ji'}^{tk'})-\frac{1}{n_j^{tk}n_j^{c}}\mathrm{Cov}(\sum_iy_{ji}^{tk},\sum_{i'}y_{ji'}^{c})\\
 &-\frac{1}{n_j^{tk'}n_j^{c}}\mathrm{Cov}(\sum_iy_{ji}^{c},\sum_{i'}y_{ji'}^{tk'})+\frac{1}{n_j^{c}n_j^{c}}\mathrm{Cov}(\sum_iy_{ji}^{c},\sum_{i'}y_{ji'}^{c})\\
 =&\frac{1}{n_j^{tk}n_j^{tk'}}\sum_{i,i'}\mathrm{Cov}(y_{ji}^{tk},y_{ji'}^{tk'})-\frac{1}{n_j^{tk}n_j^{c}}\sum_{i,i'}\mathrm{Cov}(y_{ji}^{tk},y_{ji'}^{c})\\
 &-\frac{1}{n_j^{tk'}n_j^{c}}\sum_{i,i'}\mathrm{Cov}(y_{ji}^{c},y_{ji'}^{tk'})+\frac{1}{(n_j^c)^2}\sum_{i=i'}\mathrm{Cov}(\sum_iy_{ji}^{c},\sum_{i'}y_{ji'}^{c})\\
 =&\frac{1}{(n_j^c)^2}\sum_i\mathrm{Var}(y_{ji}^c)
 \end{split}
 \end{equation}

Furthermore, if we assume the individuals in treatment groups and control group within the same study are normally distributed with the same study specific variance
 \begin{equation}
  y_{ji}^{tk}\sim N(\mu_j^{tk},\sigma_j^2)
 \end{equation}

  \begin{equation}
  y_{ji}^{c}\sim N(\mu_j^{c},\sigma_j^2)
 \end{equation}
then by derivations above, the covariance is
 \begin{equation}
 \mathrm{Cov}(MD_j^k,MD_j^{k'})
 =\frac{1}{(n_j^c)^2}\sum_i\mathrm{Var}(y_{ji}^c)
 =\frac{1}{n_j^c}\sigma_j^2
 \end{equation}

By the normal assumptions of the individuals within a study, we can also have the distribution of mean differences between treatment group $k$ and the control group
 \begin{equation}
   \overline{y}_j^{tk}-\overline{y}_j^{c}\sim N(\mu_j^{tk}-\mu_j^{c},\left(\frac{1}{n_j^{tk}}+\frac{1}{n_j^{c}}\right)\sigma_j^2)
 \label{dismd}
 \end{equation}

The pooled standard deviation of study $j$ is given by
\begin{equation}
s_{jp}=\sqrt{\frac{(n_j^{t1}-1)(s_j^{t1})^2+\cdots+(n_j^{tk_j}-1)(s_j^{tk_j})^2+(n_j^{c}-1)(s_j^{c})^2}{n_j^{t1}+\cdots+n_j^{tk_j}+n_j^{c}-(k_j+1)}}
\end{equation}
and $s_{jp}$ has the distribution
\begin{equation}
s_{jp}^2\cdot\frac{n_j^{t1}+\cdots+n_j^{tk_j}+n_j^{c}-(k_j+1)}{\sigma_j^2}\sim\chi^2_{n_j^{t1}+\cdots+n_j^{tk_j}+n_j^{c}-(k_j+1)}
\label{dissp}
\end{equation}
If we denote
\begin{equation}
v_j=n_j^{t1}+\cdots+n_j^{tk_j}+n_j^{c}-(k_j+1)
\end{equation}
then the distribution of $s_{jp}$ has a simplified form
\begin{equation}
s_{jp}^2\cdot\frac{v_j}{\sigma_j^2}\sim\chi^2_{v_j}
\end{equation}

The standardized mean difference between treatment group $k$ and control group of study $j$ is
\begin{equation}
g_j^k=J(v_j)\cdot\frac{\overline{y}_j^{tk}-\overline{y}_j^{c}}{s_{jp}}
\end{equation}

\subsection{Estimating covariance between standardized mean differences by Wei's method}
To find out $\mathrm{Cov}(g_j^k,g_j^{k'})$, we follow Wei's method in this section by applying delta method and the property of moments to get an approximation. If we denote

\begin{equation}
X=(\overline{y}_j^{tk}-\overline{y}_j^{c},\frac{1}{s_{jp}},\overline{y}_j^{tk'}-\overline{y}_j^{c})=(X_1,X_2,X_3)
\end{equation}
and
 \begin{equation}
 \begin{cases}
 f_1(X)=J(v_j)\cdot\frac{\overline{y}_j^{tk}-\overline{y}_j^{c}}{s_{jp}}=J(v_j)X_1X_2\\
 f_2(X)=J(v_j)\cdot\frac{\overline{y}_j^{tk'}-\overline{y}_j^{c}}{s_{jp}}=J(v_j)X_2X_3
 \end{cases}
\end{equation}
then by definition,
\begin{equation}
\mathrm{Cov}(g_j^k,g_j^{k'})=\mathrm{Cov}(J(v_j)\cdot\frac{\overline{y}_j^{tk}-\overline{y}_j^{c}}{s_{jp}},J(v_j)\cdot\frac{\overline{y}_j^{tk'}-\overline{y}_j^{c}}{s_{jp}})=\mathrm{Cov}(f_1(X),f_2(X))
\end{equation}

Delta method gives
 \begin{equation}
\mathrm{Var}(f(X))\approx \nabla f^T \mathrm{Var}(X)\nabla f
 \end{equation}
where
\begin{equation}
  \nabla f=
  \begin{bmatrix}
\frac{\partial f_1}{\partial X_1} & \frac{\partial f_2}{\partial X_1} \\
\frac{\partial f_1}{\partial X_2} & \frac{\partial f_2}{\partial X_2} \\
\frac{\partial f_1}{\partial X_3} & \frac{\partial f_2}{\partial X_3}
\end{bmatrix}
=J(v_j)
  \begin{bmatrix}
X_2 & 0\\
X_1 &  X_3 \\
0 & X_2
\end{bmatrix}
\end{equation}

Then we need to find out $\mathrm{E}(X)$ and $\mathrm{Var}(X)$ to evaluate $\nabla f$. By the distribution of mean differences (\ref{dismd}) we have

\begin{equation}
\begin{split}
&\mathrm{E}(MD_j^k)=\mathrm{E}(\overline{y}_j^{tk}-\overline{y}_j^{c})=\mu_j^{tk}-\mu_j^{c}\\
&\mathrm{Var}(MD_j^k)=\mathrm{Var}(\overline{y}_j^{tk}-\overline{y}_j^{c})=\left(\frac{1}{n_j^{tk}}+\frac{1}{n_j^{c}}\right)\sigma_j^2
\end{split}
\end{equation}

\begin{equation}
\begin{split}
&\mathrm{E}(MD_j^{k'})=\mathrm{E}(\overline{y}_j^{tk'}-\overline{y}_j^{c})=\mu_j^{tk'}-\mu_j^{c}\\
&\mathrm{Var}(MD_j^{k'})=\mathrm{Var}(\overline{y}_j^{tk'}-\overline{y}_j^{c})=\left(\frac{1}{n_j^{tk'}}+\frac{1}{n_j^{c}}\right)\sigma_j^2
\end{split}
\end{equation}

By (\ref{dissp}) and the moments of Chi-square distribution, we have

\begin{equation}
\mathrm{E}\left((s_{jp}^2\cdot\frac{v_j}{\sigma_j^2})^{-\frac{1}{2}}\right)=\mathrm{E}\left(\frac{1}{s_{jp}}\frac{\sigma_j}{\sqrt{v_j}}\right)=2^{-\frac{1}{2}}\cdot\frac{\Gamma(-\frac{1}{2}+\frac{v_j}{2})}{\Gamma(\frac{v_j}{2})}
\end{equation}

\begin{equation}
\mathrm{E}\left((s_{jp}^2\cdot\frac{v_j}{\sigma_j^2})^{-1}\right)=\mathrm{E}\left(\frac{\sigma_j^2}{s_{jp}^2v_j}\right)=2^{-1}\cdot\frac{\Gamma(-1+\frac{v_j}{2})}{\Gamma(\frac{v_j}{2})}
\end{equation}
divide both sides by some constants, we have
\begin{equation}
\mathrm{E}\left(\frac{1}{s_{jp}}\right)=\frac{\sqrt{\frac{v_j}{2}}\Gamma(\frac{v_j-1}{2})}{\sigma_j\Gamma(\frac{v_j}{2})}=\frac{1}{\sigma_j\cdot J(v_j)}
\end{equation}

\begin{equation}
\mathrm{E}\left(\frac{1}{s_{jp}^2}\right)=\frac{v_j}{2}\cdot\frac{1}{\sigma_j^2}\cdot\frac{\Gamma(\frac{v_j}{2}-1)}{(\frac{v_j}{2}-1)\Gamma(\frac{v_j}{2}-1)}=\frac{1}{\sigma_j^2}\cdot\frac{v_j}{v_j-2}
\end{equation}
and thus the variance is
\begin{equation}
\mathrm{Var}\left(\frac{1}{s_{jp}}\right)=\mathrm{E}\left(\frac{1}{s^2_{jp}}\right)-\left[\mathrm{E}\left(\frac{1}{s_{jp}}\right)\right]^2=\frac{1}{\sigma_j^2}\left(\frac{v_j}{v_j-2}-\frac{1}{(J(v_j))^2}\right)
\end{equation}

Then $\nabla f$ evaluated at $\mathrm{E}(X)$ is
\begin{equation}
  \nabla f=
J(v_j)
\begin{bmatrix}
\mathrm{E}(X_2) & 0\\
\mathrm{E}(X_1) & \mathrm{E}(X_3) \\
0 & \mathrm{E}(X_2)
\end{bmatrix}
=J(v_j)
\begin{bmatrix}
\frac{1}{\sigma_j\cdot J(v_j)} & 0\\
\mu_j^{tk}-\mu_j^{c} & \mu_j^{tk'}-\mu_j^{c} \\
0 & \frac{1}{\sigma_j\cdot J(v_j)}
\end{bmatrix}
\end{equation}
and we also have the variance-covariance matrix $\mathrm{Var}(X)$

\begin{equation}
\begin{split}
\mathrm{Var}(X)&=
\begin{bmatrix}
\mathrm{Cov}(X_1,X_1) & \mathrm{Cov}(X_1,X_2) &\mathrm{Cov}(X_1,X_3)\\
\mathrm{Cov}(X_1,X_2) & \mathrm{Cov}(X_2,X_2) &\mathrm{Cov}(X_2,X_3)\\
\mathrm{Cov}(X_1,X_3) & \mathrm{Cov}(X_2,X_3) &\mathrm{Cov}(X_3,X_3)
\end{bmatrix}\\
&=
\begin{bmatrix}
\left(\frac{1}{n_j^{tk}}+\frac{1}{n_j^{c}}\right)\sigma_j^2 & 0 &\frac{\sigma_j^2}{n_j^c}\\
0 & \frac{1}{\sigma_j^2}\left(\frac{v_j}{v_j-2}-\frac{1}{(J(v_j))^2}\right) & 0\\
\frac{\sigma_j^2}{n_j^c}& 0 & \left(\frac{1}{n_j^{tk'}}+\frac{1}{n_j^{c}}\right)\sigma_j^2
\end{bmatrix}
\end{split}
\end{equation}

By delta method, $\mathrm{Var}(f(X))$ can be approximated by
 \begin{equation}
 \begin{split}
&\mathrm{Var}(f(X)) \approx \nabla f^T  \mathrm{Var}(X)\nabla f\\
&=J^2(v_j)
\begin{bmatrix}
 (\mathrm{E}X_2)^2\mathrm{Var}(X_1)+(\mathrm{E}X_1)^2\mathrm{Var}(X_2)  & \mathrm{E}X_1\mathrm{E}X_3\mathrm{Var}(X_2)+(\mathrm{E}X_2)^2\mathrm{Cov}(X_1,X_3)\\
\mathrm{E}X_1\mathrm{E}X_3\mathrm{Var}(X_2)+(\mathrm{E}X_2)^2\mathrm{Cov}(X_1,X_3) & (\mathrm{E}X_3)^2\mathrm{Var}(X_2)+(\mathrm{E}X_2)^2\mathrm{Var}(X_3)
\end{bmatrix}\\
&=J^2(v_j)\\
&\begin{bmatrix}
 \frac{\frac{1}{n_j^{tk}}+\frac{1}{n_j^{c}}}{(J(v_j))^2}+\frac{(\mu_j^{tk}-\mu_j^{c})^2}{\sigma_j^2}\left(\frac{v_j}{v_j-2}-\frac{1}{(J(v_j))^2}\right)  & \frac{(\mu_j^{tk}-\mu_j^{c})(\mu_j^{tk'}-\mu_j^{c})}{\sigma_j^2}\left(\frac{v_j}{v_j-2}-\frac{1}{(J(v_j))^2}\right)+\frac{1}{n_j^c J(v_j)^2}\\
 \frac{(\mu_j^{tk}-\mu_j^{c})(\mu_j^{tk'}-\mu_j^{c})}{\sigma_j^2}\left(\frac{v_j}{v_j-2}-\frac{1}{(J(v_j))^2}\right)+\frac{1}{n_j^c J(v_j)^2} & \frac{(\mu_j^{tk'}-\mu_j^{c})^2}{\sigma_j^2}\left(\frac{v_j}{v_j-2}-\frac{1}{(J(v_j))^2}\right)+\frac{\frac{1}{n_j^{tk'}}+\frac{1}{n_j^{c}}}{ (J(v_j))^2}
\end{bmatrix}
\end{split}
 \end{equation}

Since the definition of $\mathrm{Var}(f(X))$ is

 \begin{equation}
 \mathrm{Var}(f(X))=
 \begin{pmatrix}
 \mathrm{Cov}(f_1(X),f_1(X)) & \mathrm{Cov}(f_1(X),f_2(X))  \\
 \mathrm{Cov}(f_2(X),f_1(X)) & \mathrm{Cov}(f_2(X),f_2(X))
\end{pmatrix}
 \end{equation}

Comparing each entry of  $\mathrm{Var}(f(X))$ and $\nabla f^T  \mathrm{Var}(X)\nabla f$, we have

\begin{equation}
\begin{split}
 \mathrm{Cov}(g_j^k,g_j^{k'})& =\mathrm{Cov}(f_1(X),f_2(X))\\
 &\approx J^2(v_j)\left( \frac{(\mu_j^{tk}-\mu_j^{c})(\mu_j^{tk'}-\mu_j^{c})}{\sigma_j^2}\left(\frac{v_j}{v_j-2}-\frac{1}{(J(v_j))^2}\right)+\frac{1}{n_j^c J(v_j)^2}\right)
\end{split}
\end{equation}

\subsection{Our novel method to approximate the covariance}

In this section we apply our novel method to the scenario when there are multiple treatments for one outcome in each study. The essence of our method is constructing the functions in delta method properly to avoid the difficulties in finding out the moments of complicated variables.

We denote
\begin{equation}
X=(\overline{y}_j^{tk}-\overline{y}_j^{c},s^2_{jp},\overline{y}_j^{tk'}-\overline{y}_j^{c})=(X_1,X_2,X_3)
\end{equation}
next we construct two functions
 \begin{equation}
 \begin{cases}
 f_1(X)=\frac{\overline{y}_j^{tk}-\overline{y}_j^{c}}{\sqrt{s^2_{jp}}}=\frac{X_1}{\sqrt{X_2}}\\
 f_2(X)=\frac{\overline{y}_j^{tk'}-\overline{y}_j^{c}}{\sqrt{s^2_{jp}}}=\frac{X_3}{\sqrt{X_2}}
 \end{cases}
\end{equation}
thus the covariance of standardized mean differences is
\begin{equation}
\mathrm{Cov}(g_j^k,g_j^{k'})=\mathrm{Cov}(J(v_j)\cdot\frac{\overline{y}_j^{tk}-\overline{y}_j^{c}}{s_{jp}},J(v_j)\cdot\frac{\overline{y}_j^{tk'}-\overline{y}_j^{c}}{s_{jp}})=J^2(v_j)\mathrm{Cov}(f_1(X),f_2(X))
\end{equation}

To find out $\mathrm{Cov}(g_j^k,g_j^{k'})$, we use delta method to get an approximation.
 \begin{equation}
\mathrm{Var}(f(X))\approx \nabla f^T \mathrm{Var}(X)\nabla f
 \end{equation}

where
\begin{equation}
  \nabla f=
  \begin{bmatrix}
\frac{\partial f_1}{\partial X_1} & \frac{\partial f_2}{\partial X_1} \\
\frac{\partial f_1}{\partial X_2} & \frac{\partial f_2}{\partial X_2} \\
\frac{\partial f_1}{\partial X_3} & \frac{\partial f_2}{\partial X_3}
\end{bmatrix}
=\begin{bmatrix}
\frac{1}{\sqrt{X_2}} & 0\\
-\frac{X_1}{2\sqrt{X^3_2}} & -\frac{X_3}{2\sqrt{X^3_2}} \\
0 & \frac{1}{\sqrt{X_2}}
\end{bmatrix}
\end{equation}

Then we can find out $\mathrm{E}(X)$ and $\mathrm{Var}(X)$ by (\ref{dismd}), (\ref{dissp}) and the means and variances of normal and Chi-square distributions

\begin{equation}
\begin{split}
&\mathrm{E}(X_1)=\mathrm{E}(\overline{y}_j^{tk}-\overline{y}_j^{c})=\mu_j^{tk}-\mu_j^{c}\\
&\mathrm{Var}(X_1)=\mathrm{Var}(\overline{y}_j^{tk}-\overline{y}_j^{c})=\left(\frac{1}{n_j^{tk}}+\frac{1}{n_j^{c}}\right)\sigma_j^2
\end{split}
\end{equation}

\begin{equation}
\begin{split}
&\mathrm{E}(X_3)=\mathrm{E}(\overline{y}_j^{tk'}-\overline{y}_j^{c})=\mu_j^{tk'}-\mu_j^{c}\\
&\mathrm{Var}(X_3)=\mathrm{Var}(\overline{y}_j^{tk'}-\overline{y}_j^{c})=\left(\frac{1}{n_j^{tk'}}+\frac{1}{n_j^{c}}\right)\sigma_j^2
\end{split}
\end{equation}

\begin{equation}
\mathrm{E}\left(s_{jp}^2\cdot\frac{v_j}{\sigma_j^2}\right)=v_j
\end{equation}

\begin{equation}
\mathrm{E}(X_2)=\mathrm{E}\left(s_{jp}^2\right)=\sigma_{j}^2
\end{equation}

\begin{equation}
\mathrm{Var}\left(s_{jp}^2\cdot\frac{v_j}{\sigma_j^2}\right)=2v_j
\end{equation}

\begin{equation}
\mathrm{Var}(X_2)=\mathrm{Var}\left(s_{jp}^2\right)=\frac{2\sigma_j^4}{v_j}
\end{equation}

Then $\nabla f$ evaluated at $\mathrm{E}(X)$ is
\begin{equation}
  \nabla f=
\begin{bmatrix}
\frac{1}{\sqrt{X_2}} & 0\\
-\frac{X_1}{2\sqrt{X^3_2}} & -\frac{X_3}{2\sqrt{X^3_2}} \\
0 & \frac{1}{\sqrt{X_2}}
\end{bmatrix}
=
\begin{bmatrix}
\frac{1}{\sigma_{j}} & 0\\
-\frac{\mu_j^{tk}-\mu_j^{c}}{2\sigma_{j}^3} & -\frac{\mu_j^{tk'}-\mu_j^{c}}{2\sigma_{j}^3} \\
0 & \frac{1}{\sigma_{j}}
\end{bmatrix}
\end{equation}
and we also have $\mathrm{Var}(X)$ by knowing all its entries

\begin{equation}
\begin{split}
\mathrm{Var}(X)&=
\begin{bmatrix}
\mathrm{Cov}(X_1,X_1) & \mathrm{Cov}(X_1,X_2) &\mathrm{Cov}(X_1,X_3)\\
\mathrm{Cov}(X_1,X_2) & \mathrm{Cov}(X_2,X_2) &\mathrm{Cov}(X_2,X_3)\\
\mathrm{Cov}(X_1,X_3) & \mathrm{Cov}(X_2,X_3) &\mathrm{Cov}(X_3,X_3)
\end{bmatrix}\\
&=
\begin{bmatrix}
\left(\frac{1}{n_j^{tk}}+\frac{1}{n_j^{c}}\right)\sigma_j^2 & 0 &\frac{\sigma_j^2}{n_j^c}\\
0 & \frac{2\sigma_j^4}{v_j} & 0\\
\frac{\sigma_j^2}{n_j^c}& 0 & \left(\frac{1}{n_j^{tk'}}+\frac{1}{n_j^{c}}\right)\sigma_j^2
\end{bmatrix}
\end{split}
\end{equation}

Thus

 \begin{equation}
 \begin{split}
&\nabla f^T  \mathrm{Var}(X)\nabla f\\
&=\begin{bmatrix}
 (\frac{\partial f_1}{\partial X_1})^2\mathrm{Var}(X_1)+(\frac{\partial f_1}{\partial X_2})^2\mathrm{Var}(X_2)  &
 \frac{\partial f_1}{\partial X_2}\frac{\partial f_2}{\partial X_2}\mathrm{Var}(X_2)+\frac{\partial f_1}{\partial X_1}\frac{\partial f_2}{\partial X_3}\mathrm{Cov}(X_1,X_3)\\
 \frac{\partial f_1}{\partial X_2}\frac{\partial f_2}{\partial X_2}\mathrm{Var}(X_2)+\frac{\partial f_1}{\partial X_1}\frac{\partial f_2}{\partial X_3}\mathrm{Cov}(X_1,X_3) & (\frac{\partial f_2}{\partial X_2})^2\mathrm{Var}(X_2)+(\frac{\partial f_2}{\partial X_3})^2\mathrm{Var}(X_3)
\end{bmatrix}\\
&=
\begin{bmatrix}
 \frac{1}{n_j^{tk}}+\frac{1}{n_j^{c}}+ \left(\frac{\mu_j^{tk}-\mu_j^{c}}{\sigma_j}\right)^2\cdot\frac{1}{2v_j} & \frac{(\mu_j^{tk}-\mu_j^{c})(\mu_j^{tk'}-\mu_j^{c})}{2\sigma_j^2v_j}+\frac{1}{n_j^c}\\
 \frac{(\mu_j^{tk}-\mu_j^{c})(\mu_j^{tk'}-\mu_j^{c})}{2\sigma_j^2v_j}+\frac{1}{n_j^c} & \frac{(\mu_j^{tk'}-\mu_j^{c})^2}{2v_j\sigma_j^2}+\frac{1}{n_j^{tk'}}+\frac{1}{n_j^{c}}
\end{bmatrix}
\end{split}
\end{equation}

By delta method
 \begin{equation}
\mathrm{Var}(f(X))\approx \nabla f^T  \mathrm{Var}(X)\nabla f
 \end{equation}
 and the definition of $\mathrm{Var}(f(X))$
 \begin{equation}
 \mathrm{Var}(f(X))=
 \begin{pmatrix}
 \mathrm{Cov}(f_1(X),f_1(X)) & \mathrm{Cov}(f_1(X),f_2(X))  \\
 \mathrm{Cov}(f_2(X),f_1(X)) & \mathrm{Cov}(f_2(X),f_2(X))
\end{pmatrix}
 \end{equation}

Comparing each entry of $\mathrm{Var}(f(X))$ and $\nabla f^T  \mathrm{Var}(X)\nabla f$, we have

\begin{equation}
\begin{split}
 \mathrm{Cov}(g_j^k,g_j^{k'})& =J^2(v_j)\mathrm{Cov}(f_1(X),f_2(X))\\
 &\approx J^2(v_j)\left(\frac{(\mu_j^{tk}-\mu_j^{c})(\mu_j^{tk'}-\mu_j^{c})}{2\sigma_j^2v_j}+\frac{1}{n_j^c}\right)
\end{split}
\end{equation}

\section{conclusion}

In this paper we propose a more general method to approximate the with-in study covariance of standardized mean differences which could be adapted to a broader range of effect sizes. The essence of the method is constructing proper functions of statistics with known distributions before applying delta method. At the same time, this paper also estimates the within-study variance and covariance matrix for standardized mean differences of multiple outcomes and multiple treatments respectively. 

\bibliographystyle{amsplain}

\end{document}